\documentclass[a4paper]{jpconf}
\usepackage[utf8x]{inputenc}
\usepackage{amsmath}
\usepackage{feynmf}
\usepackage{latexsym}
\usepackage{amssymb}
\usepackage{graphicx}
\usepackage{rotating}
\usepackage{latexsym}
\usepackage{verbatim}

 \def\be{\begin{equation}}
 \def\ee{\end{equation}}

 \def\s{\sigma}
 
 \def\sl{\ds}
 \def\ds#1{#1\kern-1ex\hbox{/}}
 \def\sla{\raise.15ex\hbox{$/$}\kern-.57em}
 
 \def\bea{\begin{eqnarray}}
 \def\eea{\end{eqnarray}}
 \newcommand{\bth}{{\bf 3}}
 \newcommand{\btw}{{\bf 2}}
 \newcommand{\bon}{{\bf 1}}
 \def\QQ{{Q_Q}}

 \def\QL{{Q_L}}
 
 \def\QHu{{Q_{H_u}}}

 \def\({\left(}
 \def\){\right)}
 \def\[{\left[}
 \def\]{\right]}

 \def\sl{\ds}
 \def\ds#1{#1\kern-1ex\hbox{/}}
 \def\sla{\raise.15ex\hbox{$/$}\kern-.57em}

\begin{document}
\title{Anomalous $U(1)$ and asymmetry}
\author{Andrea Mammarella}
\address{Physics Department, University of Rome, Tor Vergata, Via della
Ricerca Scientifica 1, 00133, Rome, Italy}
\ead{andrea.mammarella@roma2.infn.it}

\begin{abstract}
In the past years many possible extensions of the Standard Model (SM) have been
investigated. If one of this model is revealed at the LHC, we will need tools to 
distinguish it among the many others studied. One possibility to achieve this goal is to 
utilize the forward-backward asymmetry. In this paper we calculate the asymmetry for a model in which there is 
an extra $U(1)$ anomalous symmetry. Furthermore, we show that the asymmetry can be used to 
impose constraints on the free parameters of the model. 
\end{abstract}

\section{Introduction}
String theory results suggest that extra $U(1)$ models \cite{Langacker:2008yv} can be strong 
candidates to extend the Standard Model (SM). This led us to develop the MiAUMSSM 
\cite{Anastasopoulos:2008jt}, in which the anomalies
related to the extra $U(1)$ are cancelled by the Green-Schwarz (GS) mechanism, thus letting the 
new charges unconstrained. An alternative version of this model which admits
spontaneous supersymmetry breaking can be found in \cite{Lionetto:2011jp}.\\
We have already studied the phenomelogy of this model using cosmological
constraints, such as WMAP data on Dark Matter \cite{Fucito:2008ai,Fucito:2011kn}. We have also 
studied the radiative decays of the Next to Lightest Supersymmetric Particle
in this model \cite{Lionetto:2009dp,Fucito:2010dj}.\\
In this paper we aim to use the asymmetry to investigate the detailed properties
of the $Z'$, the new gauge boson associated to the extra $U(1)$ by fixing the charges. We also 
develop a method that, given the experimental data, can confirm or discard the model
in consideration. This method is based on the use of different definitions of asymmetry at the 
LHC.\\
At the LHC \cite{LHC} cuts on the parameter
space have to be performed. Each possible cut leads to a different definition
of asymmetry at the LHC. In the paper we choose some of these definitions to 
make our calculations. We will add further details on their explicit form in the
related section.\\
The article is organized as follow: in sec. \ref{sec2} we briefly review the properties of
the model that we are going to use, in sec. \ref{sec3} we discuss the four definitions of
asymmetry that will appear in the following and the related calculations. 
In sec. \ref{sec4} we show our results.

\section{MiAUMSSM \label{sec2}}
Our model \cite{Anastasopoulos:2008jt} is an extension of the MSSM with an extra $U(1)$ gauge symmetry.
The charges of the matter fields with respect to the symmetry gauge groups are given in table 1.\\
Gauge invariance of the model implies that only three of the $U(1)$ charges
are independent. 
We have chosen $\QQ$, $\QL$ and $\QHu$ without losing generality.
 The anomalies
induced by this extension are cancelled by the GS mechanism, so there are 
not further constraints on the charges.\\
We are going to evaluate the asymmetry of the process $pp\rightarrow e^+e^-$.
  \begin{table}[h]
  \centering
  \begin{tabular}[h]{|c||c|c|c|c|}
   \hline & SU(3)$_c$ & SU(2)$_L$  & U(1)$_Y$ & ~U(1)$^{\prime}~$\\
   \hline \hline $Q_i$   & $\bth$       &  $\btw$       &  $1/6$   & $Q_{Q}$ \\
   \hline $U^c_i$   & $\bar \bth$  &  $\bon$       &  $-2/3$  & $Q_{U^c}$
\\
   \hline $D^c_i$   & $\bar \bth$  &  $\bon$       &  $1/3$   & $Q_{D^c}$
\\
   \hline $L_i$   & $\bon$       &  $\btw$       &  $-1/2$  & $Q_{L}$ \\
   \hline $E^c_i$   & $\bon$       &  $\bon$       &  $1$     &
$Q_{E^c}$\\
   \hline $H_u$ & $\bon$       &  $\btw$       &  $1/2$   & $Q_{H_u}$\\
   \hline $H_d$ & $\bon$       &  $\btw$       &  $-1/2$  & $Q_{H_d}$ \\
   \hline
  \end{tabular}
  \caption{Charge assignment.}\label{QTable}
  \end{table}
\section{Four asymmetries at the LHC \label{sec3}}
The initial LHC state ($pp$) is symmetric. This implies that the total 
asymmetry at the LHC is zero if we integrate over the whole parameter
space. However the partonic subprocess $q\bar{q}\rightarrow e^+e^-$
is asymmetric. To keep this asymmetry we have to impose  kinematical cuts.
Each possible way to perform these cuts leads to a different definition 
of asymmetry at the LHC.\\
In this work we use the four definitions of asymmetry described in \cite{Zhou:2011dg}:
\begin{equation}
A_{\rm{RFB}}(Y_{{e\bar e}}^{\rm{cut}})=\left. \frac{\sigma(|Y_{
e}|>|Y_{{\bar e}}|)-\sigma(|Y_{ e}|<|Y_{{\bar e}}|)}{\sigma(|Y_{
e}|>|Y_{{\bar e}}|)+\sigma(|Y_{e}|<|Y_{{\bar
e}}|)}\right|_{|Y_{{e\bar e}}|>Y_{{e\bar e}}^{{cut}}}
\label{arfb} \end{equation}
\begin{equation}
A_{\rm{OFB}}(p^{\rm{cut}}_{Z,{e\bar e}})=\left. \frac{\sigma(|Y_{
e}|>|Y_{{\bar e}}|)-\sigma(|Y_{e}|<|Y_{{\bar e}}|)}{\sigma(|Y_{
e}|>|Y_{{\bar e}}|)+\sigma(|Y_{e}|<|Y_{{\bar
e}}|)}\right|_{|p_{z,{e\bar e}}|>p^{\rm{cut}}_{Z,{e\bar e}}},
\label{ao} \end{equation}
\begin{equation}
A_{\rm C}(Y_{\rm C}) = \frac{\sigma_{ e}(|Y_{e}|<Y_{\rm
C})-\sigma_{{\bar e}}(|Y_{{\bar e}}|<Y_{\rm C})} {\sigma_{
e}(|Y_{e}|<Y_{\rm C})+\sigma_{{\bar e}}(|Y_{{\bar e}}|<Y_{\rm C})}
\label{ac} \end{equation}
\begin{equation}
A_{\rm E}(Y_{\rm C}) = \frac{\sigma_{e}(Y_{\rm C}<|Y_{
e}|)-\sigma_{{\bar e}}(Y_{\rm C}<|Y_{{\bar e}}|)} {\sigma_{
e}(Y_{\rm C}<|Y_{e}|)+\sigma_{{\bar e}}(Y_{\rm C}<|Y_{{\bar e}}|)}
\label{ae} \end{equation} The first two asymmetries are defined in the center of mass 
(CM) frame. The forward-backward asymmetry $A_{RFB}$ 
\cite{Langacker:2008yv,Langacker:1984dc,Petriello:2008zr,Cvetic:1995zs,Dittmar:2003ir,Godfrey:2008vf} 
has a cut on the rapidity $Y$ of the $f/\bar{f}$ pair (in this work
$f$ is an electron), while the one side asymmetry 
$A_{O}$ \cite{Wang:2010tg,Wang:2010du} has a cut on $p_{z,f\bar{f}}$, the longitudinal component of the fermionic
pair momentum. The other two asymmetries are defined in the lab frame.
The variable $Y_C$ is the pseudo-rapidity $-\log(tan(\theta/2))$: in this case the kinematical cut is
on the angles of the outgoing particles with respect to
the beam direction in the lab frame. The central asymmetry 
$A_C$ \cite{Ferrario:2009ns,Kuhn:1998jr,Kuhn:1998kw,Antunano:2007da,Ferrario:2008wm} is
calculated integrating in the angular region centered on the axis othogonal
to the beam, while the edge asymmetry $A_E$ \cite{Xiao:2011kp} is defined in the complementar region.\\
\section{Calculations and results \label{sec4}}
We calculate the value of the cuts that maximizes the significance of each definition of asymmetry, then
we use these best cuts to evaluate the asymmetries in different ways, i.e. as a functions of two 
free charges keeping the third fixed to $0$ and as a function of the three charges. We have set
the mass of the $Z'$ boson at $1.5~TeV$ to permit a direct comparison with the results of \cite{Zhou:2011dg}.
\subsection{Optimal cuts}
\begin{figure}[h!]
\centering
 \includegraphics[scale=0.65]{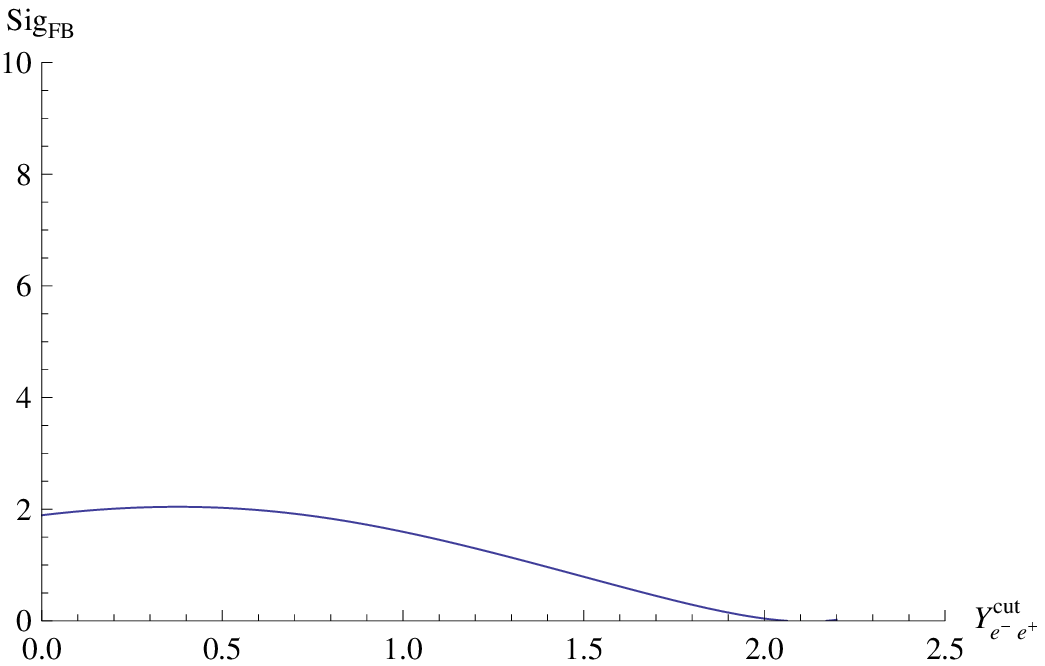}
 \includegraphics[scale=0.65]{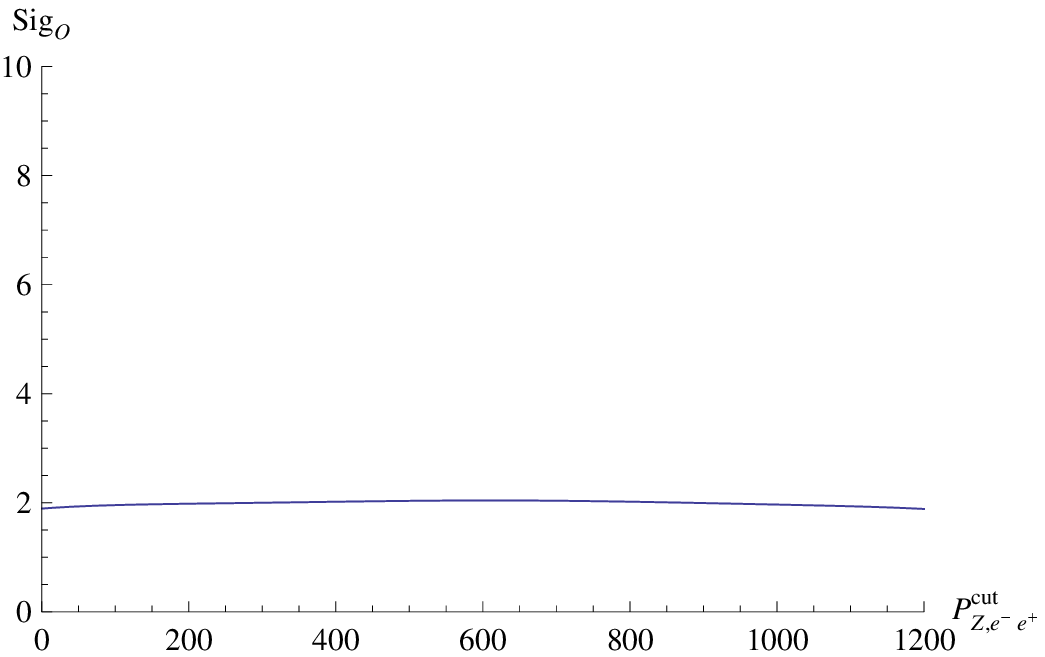}
 \includegraphics[scale=0.65]{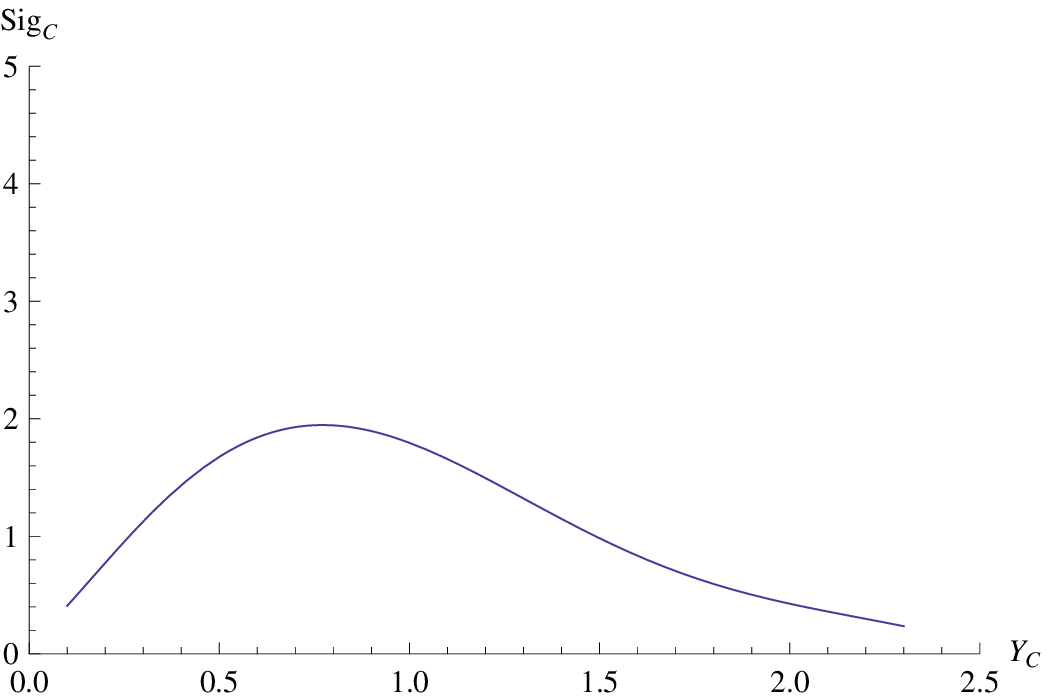}
 \includegraphics[scale=0.65]{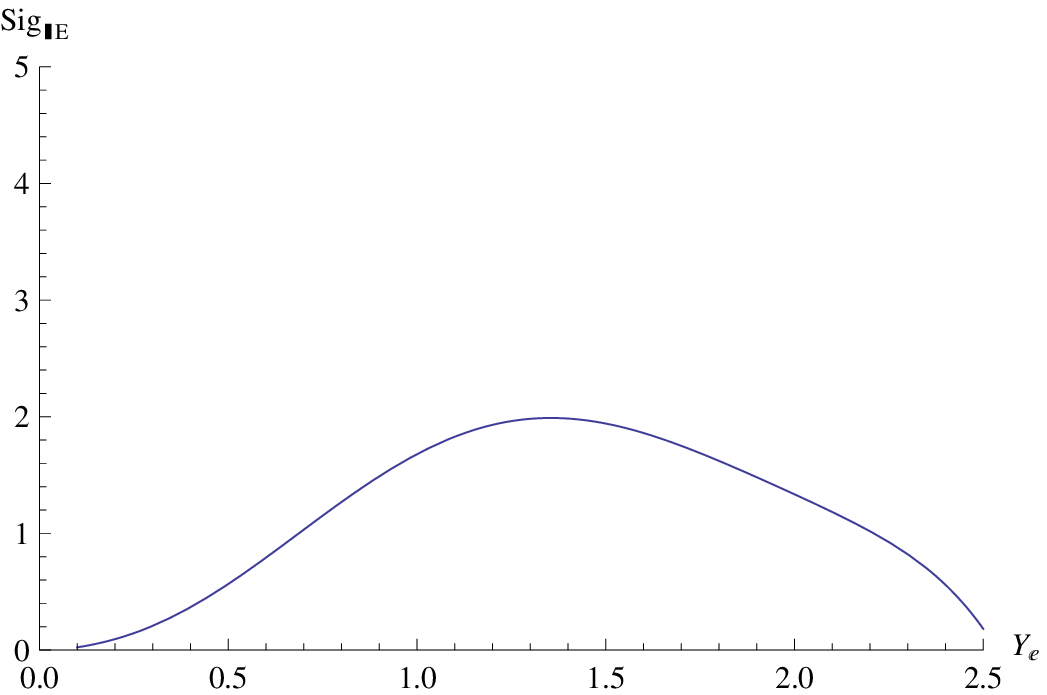}
\caption{Significance of the four definitions of asymmetry as a function 
of the corresponding cut for for on-peak events}
\label{4sig}
\end{figure}

As shown in \cite{Zhou:2011dg} the asymmetry magnitude is not a good
function to optimize the observable. A better choice is, instead, the 
statistical significance:

\be Sig \equiv A \sqrt{\mathcal{L} \sigma} \ee where $A$ can be any of the 
previously defined asymmetries, $\mathcal{L}$ is the LHC integrated
luminosity, which we take to be $100~fb^{-1}$ and $\s$ is the total 
cross section.\\
In figure \ref{4sig} we show the results associated to the significance for
all the asymmetry definition that we use in the \textquotedblleft on peak \textquotedblright
region, that is $M_{Z'}-3 \Gamma_{Z'}<M_{e^+e^-}<M_{Z'}+3 \Gamma_{Z'}$.\\
 The best cuts are those that maximize the significance.
We list them in the table \ref{bestcuts}

\begin{table}[h]
 \centering
\begin{tabular}[h]{|c||c|c|c|c|}
 \hline & $A_{RFB}$ & $A_{O}$ & $A_C$ & $A_E$  \\
 \hline \hline Best cut & $Y_{f\bar{f}}^{cut}=0.4$ & $p_{z,f\bar{f}}^{cut}=580~ GeV$ & $Y_C=0.8$ & $Y_C=1.4$ \\
 \hline 
\end{tabular}
\caption{Best cuts for
the on-peak $e^+e^-$ asymmetries}
\label{bestcuts}
\end{table}
\subsection{Asymmetry calculations}
\begin{figure}
\centering
 \includegraphics[scale=0.5]{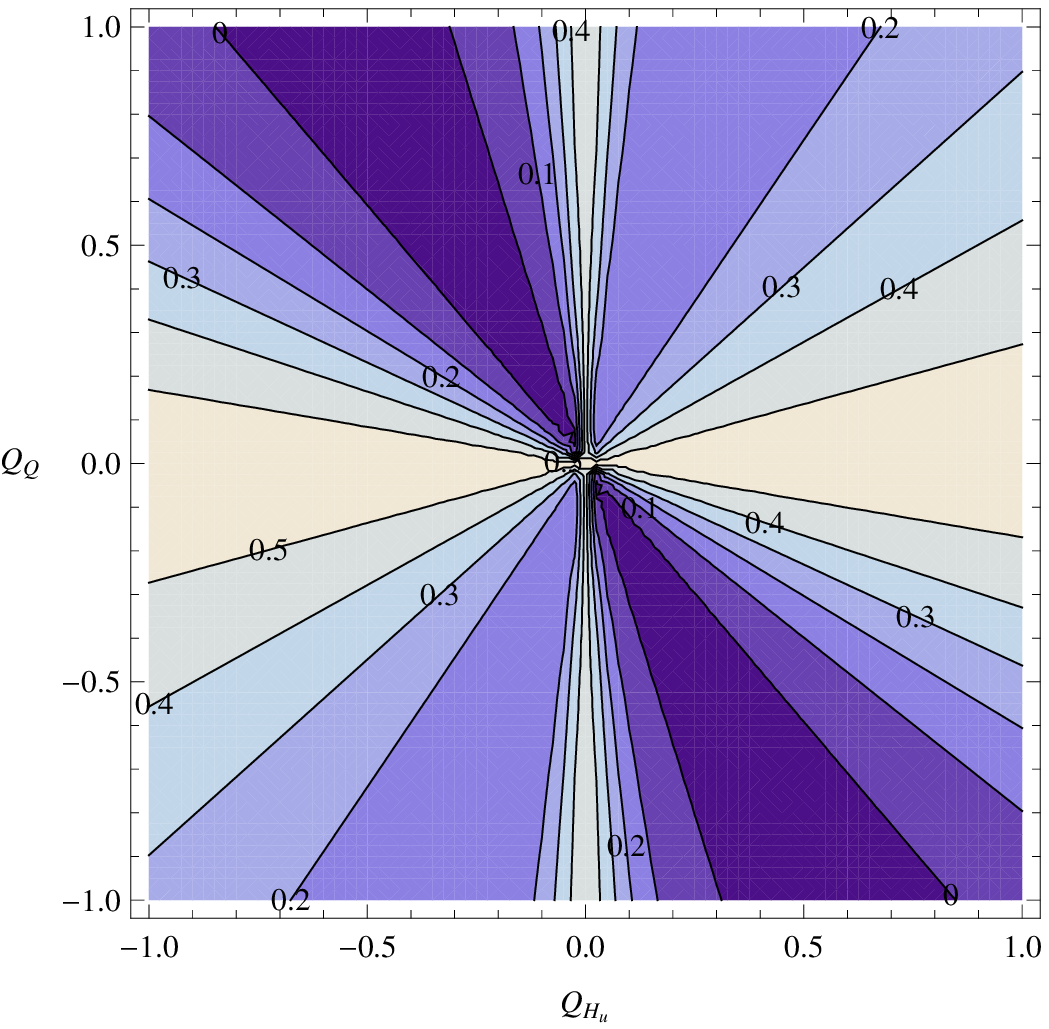}
 \includegraphics[scale=0.5]{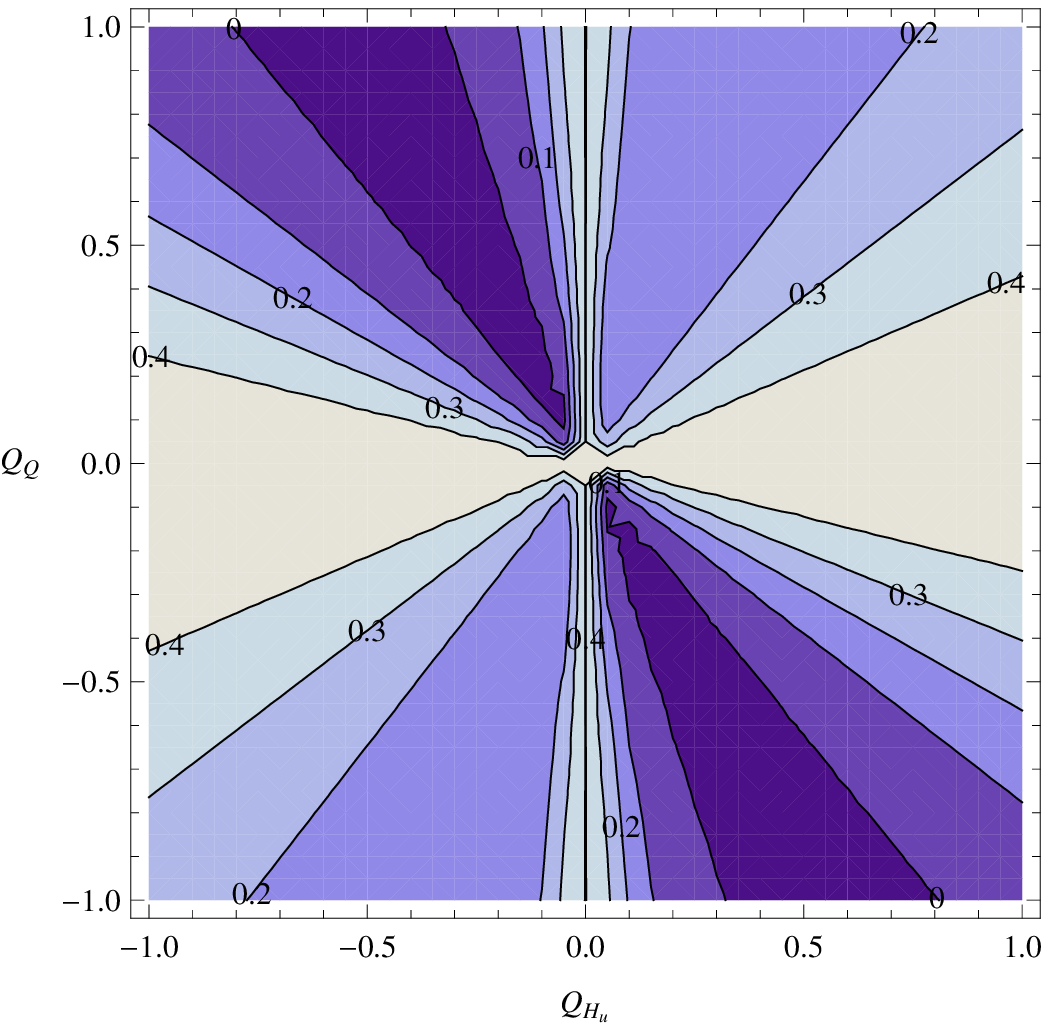}
\caption{Results with the asymmetries for $Q_{L}=0$ for the best cuts. The 
left image is $A_{RFB}$, the right image is $A_O$}
\label{a1cp}
\end{figure}

\begin{figure}
\centering
 \includegraphics[scale=0.5]{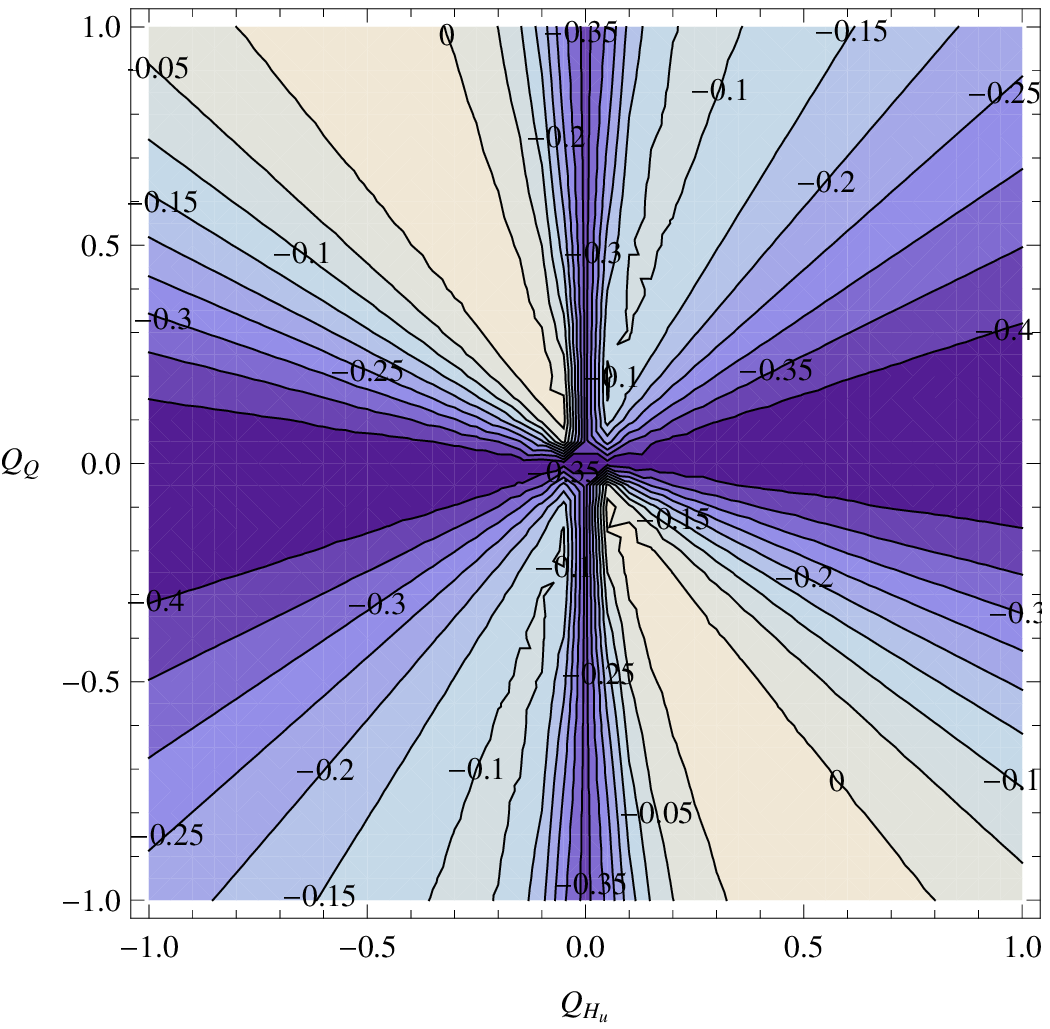}
 \includegraphics[scale=0.5]{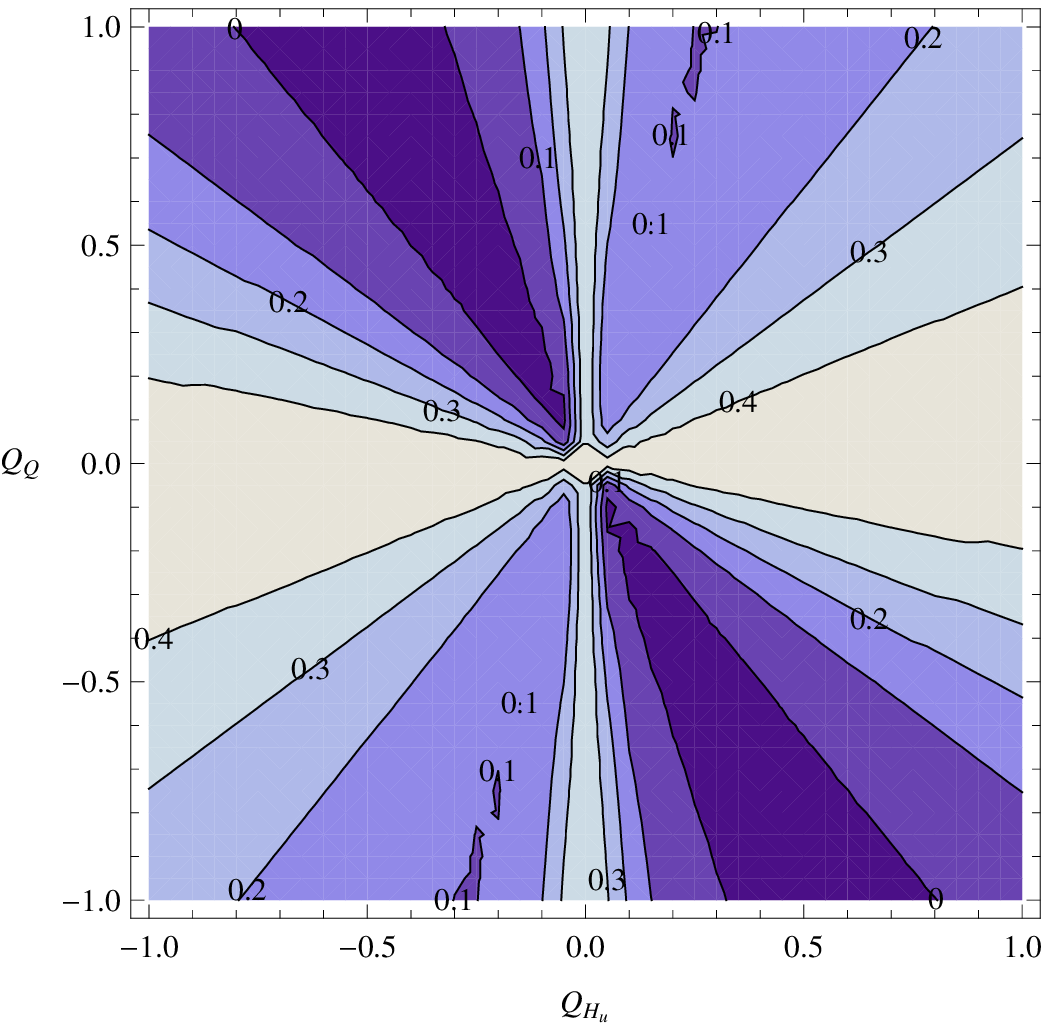}
\caption{Results for the asymmetries with $Q_{L}=0$ for the best cuts. The 
left image is $A_C$, the right image is $A_E$}
\label{a1cp1}
\end{figure}

Using the best cuts we have studied the dependence of the asymmetry from couples of the free
charges. There are three possibilities for the couples of the charges: $Q_{H_u}-Q_L$ ($Q_Q=0$), 
$Q_{H_u}-Q_Q$ ($Q_L=0$) and $Q_L-Q_Q$ ($Q_{H_u}=0$). For the sake of brevity, in figures 
\ref{a1cp} and \ref{a1cp1} we show only the plots obtained keeping $Q_L=0$.

\subsection{Charges constraints}
To impose constraints on the charges, we have calculated the asymmetry dependence from the three charges.
It turns out that the asymmetries are rational functions of 4th order in the charges, so we
use the following fit function:
  \be
 A=\frac{\sum_{i,j,k=0}^n a_{ijk}(Q_{H_u})^i(Q_Q)^j(Q_L)^k}{\sum_{i,j,k=0}^n b_{ijk}(Q_{H_u})^i(Q_Q)^j(Q_L)^k}
\label{AQi}
\ee with $i+j+k= n\le 4$.
We have calculated this expansion for the previous four asymmetry definitions that we have considered 
so we have them expressed as a functions of the charges. We do not show here their complete expression, 
that can be found in %\cite{}%
However, if a deviation from the asymmetry associated to the SM asymmetry is discovered, 
this expressions can be used as a system of four equation of the three variables $Q_{H_u} $, $Q_L$
and $Q_Q$. Solving this system using the experimental values of the asymmetry will give constraints on 
the charges and a consistency check of the model.
\section{Conclusion}
We have studied four different definitions of asymmetry at the LHC, namely the forward-backward, the one-side, 
the central and the edge asymmetries in the channel $pp\rightarrow e^+e^-$.\\
To perform this study we have written a numerical code that finds the value of the cut that maximizes the 
significance of each definition of asymmetry. After that, we have calculated the value of the four asymmetries
as functions of couple of charges, keeping one of the three free charges of our model fixed to $0$.\\
Finally, we have calculated the functional dependence of the asymmetries from the three charges using an
expansion in rational functions. If an asymmetry of this type will be found at the LHC this expansions could
give constraints on the free charges of the model.\\
So we have completely characterized the asymmetry of the MiAUSSM with a procedure that can be replied for every
other model of physics beyond the SM.


\section*{References}
\begin{thebibliography}{100}
 %\cite{Langacker:2008yv}
\bibitem{Langacker:2008yv}
  P.~Langacker,
  ``The Physics of Heavy $Z^\prime$ Gauge Bosons'',
  Rev.\ Mod.\ Phys.\  {\bf 81} (2009) 1199
  [arXiv:0801.1345 [hep-ph]].
  %%CITATION = RMPHA,81,1199;%%

%\cite{Anastasopoulos:2008jt}
\bibitem{Anastasopoulos:2008jt}
  P.~Anastasopoulos, F.~Fucito, A.~Lionetto, G.~Pradisi, A.~Racioppi and Y.~S.~Stanev,
  ``Minimal Anomalous U(1)-prime Extension of the MSSM'',
  Phys.\ Rev.\  D {\bf 78} (2008) 085014
  [arXiv:0804.1156 [hep-th]].
  %%CITATION = PHRVA,D78,085014;%%

%\cite{Lionetto:2011jp}
\bibitem{Lionetto:2011jp}
  A.~Lionetto and A.~Racioppi,
  ``Supersymmetry Breaking in a Minimal Anomalous Extension of the MSSM'',
  arXiv:1102.5040 [hep-ph].
  %%CITATION = ARXIV:1102.5040;%%

%\cite{Fucito:2008ai}
\bibitem{Fucito:2008ai}
  F.~Fucito, A.~Lionetto, A.~Mammarella and A.~Racioppi,
  ``Stueckelino dark matter in anomalous U(1)-prime models'',
  Eur.\ Phys.\ J.\  C {\bf 69} (2010) 455
  [arXiv:0811.1953 [hep-ph]].
  %%CITATION = EPHJA,C69,455;%%

%\cite{Fucito:2011kn}
\bibitem{Fucito:2011kn}
  F.~Fucito, A.~Lionetto and A.~Mammarella,
  %``Dark Matter in Anomalous $U(1)'$ Models with neutral mixing,''
  Phys.\ Rev.\  D {\bf 84} (2011) 051702
  [arXiv:1105.4753 [hep-ph]].
  %%CITATION = PHRVA,D84,051702;%%

%\cite{Lionetto:2009dp}
\bibitem{Lionetto:2009dp}
  A.~Lionetto and A.~Racioppi,
  ``Gaugino radiative decay in an anomalous U(1)-prime model'',
  Nucl.\ Phys.\  B {\bf 831} (2010) 329
  [arXiv:0905.4607 [hep-ph]].
  %%CITATION = NUPHA,B831,329;%%


%\cite{Fucito:2010dj}
\bibitem{Fucito:2010dj}
  F.~Fucito, A.~Lionetto, A.~Racioppi and D.~Ricci~Pacifici,
  ``A Phenomenological study on the wino radiative decay in anomalous $U(1)'$
  models'',
  Phys.\ Rev.\  D {\bf 82} (2010) 115004
  [arXiv:1007.5443 [hep-ph]].
  %%CITATION = PHRVA,D82,115004;%%

\bibitem{LHC}
E.~Lyndon and B.~Philip (ed.), {\it LHC Machine}, JINST {\bf 3} (2008)
S08001.

%\cite{Zhou:2011dg}
\bibitem{Zhou:2011dg}
  Z.~q.~Zhou, B.~Xiao, Y.~k.~Wang and S.~h.~Zhu,
  ``Discriminating Different $Z^\prime$s via Asymmetries at the LHC'',
  Phys.\ Rev.\  D {\bf 83} (2011) 094022
  [arXiv:1102.1044 [hep-ph]].
  %%CITATION = PHRVA,D83,094022;%%

%\cite{Langacker:1984dc}
\bibitem{Langacker:1984dc}
  P.~Langacker, R.~W.~Robinett, J.~L.~Rosner,
  ``New Heavy Gauge Bosons in p p and p anti-p Collisions'',
  Phys.\ Rev.\  {\bf D30 } (1984)  1470.

%\cite{Petriello:2008zr}
\bibitem{Petriello:2008zr}
  F.~Petriello and S.~Quackenbush,
  ``Measuring $Z^\prime$ couplings at the CERN LHC'',
  Phys.\ Rev.\  D {\bf 77} (2008) 115004
  [arXiv:0801.4389 [hep-ph]].
  %%CITATION = PHRVA,D77,115004;%%

%\cite{Cvetic:1995zs}
\bibitem{Cvetic:1995zs}
  M.~Cvetic and S.~Godfrey,
  ``Discovery and identification of extra gauge bosons'',
  arXiv:hep-ph/9504216.
  %%CITATION = HEP-PH/9504216;%%

%\cite{Dittmar:2003ir}
\bibitem{Dittmar:2003ir}
  M.~Dittmar, A.~S.~Nicollerat and A.~Djouadi,
  ``Z-prime studies at the LHC: An Update'',
  Phys.\ Lett.\  B {\bf 583} (2004) 111
  [arXiv:hep-ph/0307020].
  %%CITATION = PHLTA,B583,111;%%


%\cite{Godfrey:2008vf}
\bibitem{Godfrey:2008vf}
  S.~Godfrey, T.~A.~W.~Martin,
  ``Identification of Extra Neutral Gauge Bosons at the LHC Using b- and t-Quarks'',
  Phys.\ Rev.\ Lett.\  {\bf 101}, 151803 (2008).
  [arXiv:0807.1080 [hep-ph]].

%\cite{Wang:2010tg}
\bibitem{Wang:2010tg}
  Y.~-k.~Wang, B.~Xiao, S.~-h.~Zhu,
  ``One-side forward-backward asymmetry at the LHC'',
  Phys.\ Rev.\  {\bf D83 } (2011)  015002.
  [arXiv:1011.1428 [hep-ph]].

%\cite{Wang:2010du}
\bibitem{Wang:2010du}
  Y.~-k.~Wang, B.~Xiao, S.~-h.~Zhu,
  %``One-side Forward-backward Asymmetry in Top Quark Pair Production at CERN Large Hadron Collider,''
  Phys.\ Rev.\  {\bf D82 } (2010)  094011.
  [arXiv:1008.2685 [hep-ph]].

%\cite{Ferrario:2009ns}
\bibitem{Ferrario:2009ns}
  P.~Ferrario and G.~Rodrigo,
  ``Charge asymmetries of top quarks: A Window to new physics at hadron
  colliders'',
  J.\ Phys.\ Conf.\ Ser.\  {\bf 171} (2009) 012091
  [arXiv:0907.0096 [hep-ph]].
  %%CITATION = 00462,171,012091;%%

%\cite{Kuhn:1998jr}
\bibitem{Kuhn:1998jr}
  J.~H.~Kuhn, G.~Rodrigo,
  ``Charge asymmetry in hadroproduction of heavy quarks'',
  Phys.\ Rev.\ Lett.\  {\bf 81 } (1998)  49-52.
  [hep-ph/9802268].

%\cite{Kuhn:1998kw}
\bibitem{Kuhn:1998kw}
  J.~H.~Kuhn, G.~Rodrigo,
  ``Charge asymmetry of heavy quarks at hadron colliders'',
  Phys.\ Rev.\  {\bf D59 } (1999)  054017.
  [hep-ph/9807420].

%\cite{Antunano:2007da}
\bibitem{Antunano:2007da}
  O.~Antunano, J.~H.~Kuhn, G.~Rodrigo,
  ``Top quarks, axigluons and charge asymmetries at hadron colliders'',
  Phys.\ Rev.\  {\bf D77 } (2008)  014003.
  [arXiv:0709.1652 [hep-ph]].


%\cite{Ferrario:2008wm}
\bibitem{Ferrario:2008wm}
  P.~Ferrario, G.~Rodrigo,
  ``Massive color-octet bosons and the charge asymmetries of top quarks at hadron colliders'',
  Phys.\ Rev.\  {\bf D78 } (2008)  094018.
  [arXiv:0809.3354 [hep-ph]].


%\cite{Xiao:2011kp}
\bibitem{Xiao:2011kp}
  B.~Xiao, Y.~K.~Wang, Z.~Q.~Zhou and S.~h.~Zhu,
  ``Edge Charge Asymmetry in Top Pair Production at the LHC'',
  Phys.\ Rev.\  D {\bf 83} (2011) 057503
  [arXiv:1101.2507 [hep-ph]].
  %%CITATION = PHRVA,D83,057503;%%
\end{thebibliography}
\end{document}